\documentclass[twocolumn,aps,floatfix]{revtex4}
\usepackage{amssymb}
\usepackage{graphicx}
\usepackage{amsmath}

\begin{document}
\title{Coherence properties of spinor condensates at finite temperatures}

\author{Krzysztof Gawryluk,$\,^1$ Miros{\l}aw Brewczyk,$\,^1$
        Mariusz Gajda,$\,^2$ and Kazimierz Rz{\c a}\.zewski$\,^3$ }

\affiliation{\mbox{$^1$ Instytut Fizyki Teoretycznej, Uniwersytet w Bia{\l}ymstoku,
                        ulica Lipowa 41, 15-424 Bia{\l}ystok, Poland}  \\
\mbox{$^2$ Instytut Fizyki PAN, Aleja Lotnik\'ow 32/46, 02-668 Warsaw,
           Poland}  \\
\mbox{$^3$ Centrum Fizyki Teoretycznej PAN, Aleja Lotnik\'ow 32/46, 02-668 Warsaw,
           Poland}  }

\date{\today}

\begin{abstract}
We consider a spinor condensate of $^{87}$Rb atoms in its $F=1$ hyperfine state at
finite temperatures. Putting initially all atoms in $m_F=0$ component we find that the
system evolves into the state of thermal equilibrium. This state is approached in a step-like 
process and when established it manifests itself in distinguishable ways. The atoms in states 
$m_F=+1$ and $m_F=-1$ start to rotate in opposite directions breaking the chiral symmetry
and showing highly regular spin textures. Also the coherence properties of the system changes 
dramatically. Depending on the strength of spin-changing collisions the system first enters 
the stage where the $m_F=+1$ and $m_F=-1$ spinor condensate components periodically loose 
and recover their mutual coherence whereas their thermal counterparts get completely dephased. 
For stronger spin changing collisions the system enters the regime where also the strong 
coherence between other components is built up.
\end{abstract}

\maketitle

Since a $^{23}$Na atoms Bose-Einstein condensate was created in a pure optical trap 
\cite{Ketterle} it has become possible to study the spinor properties of alkali dilute
quantum gases. In such (spinor) condensates the atoms can change their spins upon collisions 
and still remain trapped. Therefore, in spinor condensates the population of their components 
is allowed to change in time. This new degree of freedom distinguishes the spinor condensates 
from mixtures of quantum gases. 

There has been recently a growing interest in such systems. Different ground-state phases 
(ferromagnetic or polar) have been observed experimentally in $F=1$ condensates 
\cite{Ketterle1, Hamburg, Chang}. In particular, it has been shown that the ground state 
of rubidium condensate exhibits a ferromagnetic behavior whereas the sodium system reveals 
the polar phase. For a $F=2$ manifold the set of possible ground states of spinor condensates
is even richer, including the so-called cyclic phase \cite{cyclic}. Experiment proved that 
under the natural conditions a $^{87}$Rb condensate shows a polar behavior \cite{Hamburg}.
Recently, a first $F=3$ ($^{52}$Cr) condensate has been achieved \cite{Pfau} and many new 
spin phases were predicted for this system \cite{Santos}.

There has been also recently an interest in the dynamics of the spinor condensates,
including the coherent oscillations between various components \cite{Hamburg,Chang},
formation of domains \cite{domaine}, magnetically tuned resonances \cite{Bongs}, and
predicted theoretically the effect of breaking of chiral symmetry \cite{chiral}.
Some theoretical effort has been made to explore the dipolar interactions in addition
to the short-range (van der Waals) interactions. In Ref. \cite{Santos,EdH} the coupling
between the spin and orbital motion, inherently built in the dipolar interaction, is
investigated and the phenomenon of the Einstein-de Haas effect is exhibited. As has been 
shown in Ref. \cite{resonance} this effect might be observable even in a very weak dipolar
systems such as rubidium condensate provided the experiment is performed under resonance
condition.

In this Letter we investigate temperature properties of a spinor condensate of $^{87}$Rb 
atoms in $F=1$ hyperfine state. The influence of finite temperature on spinor evolution
has been studied recently experimentally \cite{Hamburg1} and theoretically although in
a quasi-one-dimensional system \cite{Maciek}.
In the second quantization notation, the Hamiltonian of the system is given by \cite{Ho}
\begin{eqnarray}
H &=& \int d^3r \left[ \hat{\psi}^{\dagger}_i(\mathbf{r}) H_0 \hat{\psi}_i(\mathbf{r})
+ \frac{c_0}{2}\, \hat{\psi}^{\dagger}_j(\mathbf{r})
\hat{\psi}^{\dagger}_i(\mathbf{r}) \hat{\psi}_i(\mathbf{r}) \hat{\psi}_j(\mathbf{r})
\right.  \nonumber  \\
&+&\left. \frac{c_2}{2}\, \hat{\psi}^{\dagger}_k(\mathbf{r})
\hat{\psi}^{\dagger}_i(\mathbf{r})\, \mathbf{F}_{ij} \mathbf{F}_{kl}\, 
\hat{\psi}_j(\mathbf{r}) \hat{\psi}_l(\mathbf{r}) \right]      \,,
\label{Hamspin}
\end{eqnarray}
where repeated indices have to be summed over the values $m_F=+1,0,-1$. The field operator 
$\hat{\psi}_i(\mathbf{r})$ annihilates an atom in the hyperfine state $|F=1,m_F=i>$ at point 
$\mathbf{r}$. The Hamiltonian consists of the kinetic energy and the trapping potential 
($H_0=-\frac{\hbar^2}{2M} \nabla^2 + V_{tr}$, where $M$ is the mass of an atom) and two terms 
which describe the spin-independent and spin-dependent parts of the contact interactions, 
respectively. Coefficients $c_0$ and $c_2$ can be expressed with the help of the scattering 
lengths $a_0$ and $a_2$ which determine the collision of atoms in a channel of total spin 
$0$ and $2$. The appropriate formulas are given by $c_0=4\pi\hbar^2 (a_0+2a_2)/3M$ and 
$c_2=4\pi\hbar^2 (a_2-a_0)/3M$ \cite{Ho}. According to Ref. \cite{a0a2}, $a_0=5.387$\,nm 
and $a_2=5.313$\,nm. Moreover, $\mathbf{F}$ are the spin-$1$ matrices.

Within the classical fields approximation \cite{przeglad} one replaces the field operators 
$\hat{\psi}_i(\mathbf{r})$ by the classical wavefunctions ${\psi}_i(\mathbf{r})$.
The equation of motion for these wavefunctions is written as
\begin{eqnarray}
i\hbar \frac{\partial}{\partial t}
\left(
\begin{array}{l}
{\psi}_1 \\
{\psi}_0 \\
{\psi}_{-1}
\end{array}
\right)
=
{\cal{H}}
\left(
\begin{array}{l}
{\psi}_1 \\
{\psi}_0 \\
{\psi}_{-1}
\end{array}
\right)   \,\,.
\label{Eqmot}
\end{eqnarray}
The diagonal part of ${\cal{H}}$ is given by
${\cal{H}}_{11} = H_0+(c_0+c_2)\, {|{\psi}_1|^2}
+(c_0+c_2)\, {|{\psi}_0|^2} + (c_0-c_2)\, {|{\psi}_{-1}|^2}  , \;
{\cal{H}}_{00} = H_0+(c_0+c_2)\, {|{\psi}_1|^2} + c_0\, {|{\psi}_0|^2}
+ (c_0+c_2)\, |{\psi}_{-1}|^2 , \;
{\cal{H}}_{-1-1} = H_0+(c_0-c_2)\, |{\psi}_1|^2 + (c_0+c_2)\, |{\psi}_0|^2
+ (c_0+c_2)\, |{\psi}_{-1}|^2 $.
The off-diagonal terms describe the collisions which do not preserve the projection of
spin of each atom (however, the total spin projection is conserved) and equal
${\cal{H}}_{10} = c_2\, {\psi}^{*}_{-1} {\psi}_0  , \;
{\cal{H}}_{0-1} = c_2\, {\psi}^{*}_0 {\psi}_1  $.
Moreover, ${\cal{H}}_{1-1} = 0$.

\begin{figure}[thb]
\begin{center} \resizebox{3.4in}{3.2in}
{\includegraphics{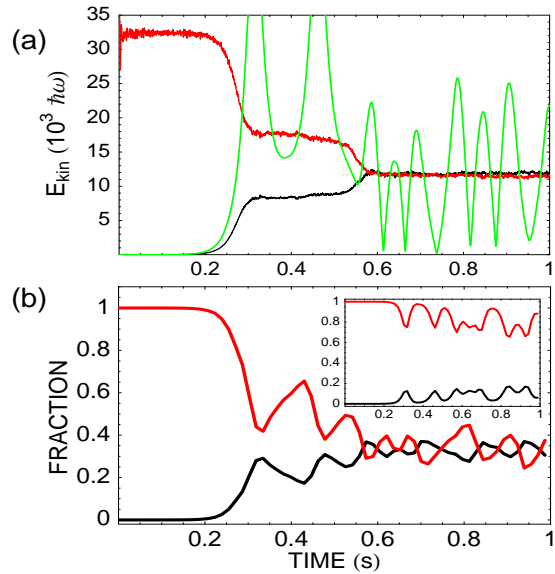}}
\caption{(color online). (a) Kinetic energy of spinor components as a function of time
for $m_F=0$ (red color) and $m_F=\pm 1$ (black color) Zeeman states. 
The green line is the overlap between the $+1$ and $-1$ components defined by the 
scalar product $|(\psi_{+1},\psi_{-1})|$.
(b) Fraction of total thermal (main frame) and total condensed (inset) atoms in each spinor 
component (red for $0$ and black for $\pm1$ states).}
\label{kin}
\end{center}
\end{figure}

To study the properties of a spinor condensate at finite temperatures we prepare initially 
an atomic sample in the $m_F=0$ component \cite{Hamburg} and allow the system to evolve. 
The Bose gas is confined in a pancake-shaped trap with the aspect ratio 
$\beta=\omega_z/\omega_r=80$ and the radial frequency $\omega_r=2\pi \times 100$\,Hz. To 
simplify calculations, we solve numerically the Eq. (\ref{Eqmot}) on a two-dimensional grid 
using rescaled scattering lengths (i.e., multiplied by a factor $\sqrt{\beta / 2\pi}$). 
Obviously, spin changing collisions allow transitions to other Zeeman states ($m_F=\pm 1$) 
and hence the interplay between the spin dynamics and the thermalization processes can be 
studied. In Fig. \ref{kin}a we plot the kinetic energy of each spinor component as a function 
of time for the initial thermal noise introduced to the $m_F=0$ component 
resulting in its $\approx 20 \%$ depletion. We find that after
some time the kinetic energy becomes equally distributed among spinor components (see Fig.
\ref{kin}), suggesting the system reached an equilibrium. The time needed for that depends 
on the initial amount of the thermal noise (the higher thermal energy the longer relaxation 
time). However, approaching this equilibrium is rather a step-like process not a uniform one. 
This is because the atomic cloud becomes fragmented and the overlap of different spin components 
changes significantly in time (Fig. \ref{kin}a, green curve). The steps in Fig. \ref{kin}a 
seem to correspond to the maximal overlap. 

Spin changing collisions between atoms in $m_F=0$ state produce equally atoms in states 
$m_F=\pm1$. Initially, both condensed and thermal atoms go to the $m_F=\pm 1$ 
components and back (see Fig. \ref{kin}b). This behavior changes after the equilibrium 
is established, i.e., after $0.6$\,s. From now on, the fraction of thermal atoms in each 
component becomes $1/3$ (Fig. \ref{kin}b, main frame). Since the total number of thermal 
atoms is also constant the system reaches the thermal equilibrium meaning all spin 
components in the thermal cloud are equally populated (already observed experimentally
\cite{Hamburg1}).
At the same time a non-trivial spin dynamics for condensed atoms is observed 
(Fig. \ref{kin}b, inset). Thus, the spinor condensate decouples from the thermal cloud. 
Also, since the condensed atoms do not populate equally spinor components (as the thermal 
atoms do) it means that the single mode approximation is inappropriate at finite temperatures.

\begin{figure}[thb]
\begin{center} \resizebox{2.9in}{1.6in}
{\includegraphics{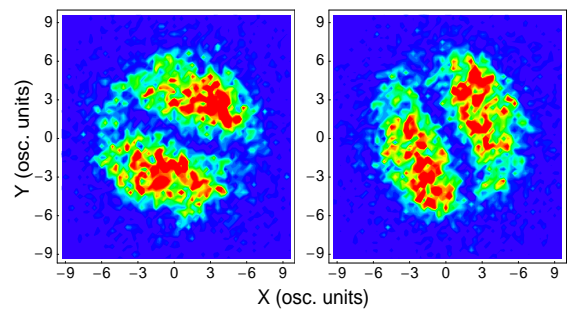}}
\caption{(color online). Density of $+1$ (left frame) and $-1$ (right frame) components 
for $c_2=-1.0\times10^{-3}$, $c_0=0.22$, and $E_{kin}(0)=9\times10^3$ osc. units 
at $0.8$\,s.   }
\label{density}
\end{center}
\end{figure}

To discuss the coherence properties of spinor condensates it is necessary to have in mind
the stability diagram for such a system. This diagram is formed as a result of
the Bogoliubov-de Gennes analysis of Eq. (\ref{Eqmot}) in two-dimensional space and is
considered with respect to the parameters $(c_2,c_0)$. More precisely, the stability of 
$m_F=0$ component against the excitations in $m_F=\pm1$ states is investigated with particular 
emphasis on modes of types $\propto e^{i m \phi}$, where $m=\pm1,0$ (see Ref. \cite{chiral}). 
This analysis proves the existence of the regions where all $m=\pm1,0$ modes are stable, 
only modes $m=\pm1$ are unstable, and all $m=\pm1,0$ modes are unstable. The coherence 
properties strongly depend on which region the systems parameters belong to. For example,
if modes $m=\pm1$ get unstable the superposition of $\propto e^{i \phi}$ and
$\propto e^{-i \phi}$ states is excited in $m_F=1$ (as well as in $m_F=-1$) component.
Assuming both modes equally contribute to this superposition one should expect the 
fragmented (and consisted of two parts) densities in $m_F=\pm1$ components. Examples
are given in Fig. \ref{density}. This fragmentation and circulation of atomic clouds
are responsible for the coherence properties discussed below. Note also that the thermal
atoms cause the granularity of densities since decreasing the level of initial thermal 
noise leads to very smooth densities.

To investigate in detail the coherence properties of spinor condensates we calculate the
one-particle density matrix by averaging the product 
$\Psi^{*}(\mathbf{r},t) \Psi(\mathbf{r'},t)$ (with the spinor classical field 
$\Psi(\mathbf{r},t)$ being the solution of Eq. (\ref{Eqmot})) over the finite volume 
\cite{przeglad}. Next, this matrix is diagonalized and an eigenvector corresponding to the 
dominant eigenvalue represents the spinor condensate wave function. Let us denote it by 
$\varphi_0(\mathbf{r},t)$. Then the thermal component of the spinor classical field can be 
defined as $\psi^T=\Psi - (\varphi_0,\Psi)\, \varphi_0$ (here, the vector $\varphi_0$ is 
normalized to unity). Thermal part of the classical field lives in a space orthogonal to 
the condensate wave function. Now we introduce the spin density matrices for the condensate 
and the thermal part as below
\begin{eqnarray}
\rho_{ij}^0 &=& \int \varphi_{0i}^*(\mathbf{r})\,  \varphi_{0j}(\mathbf{r}) \, d^3r 
\nonumber  \\
\rho_{ij}^T &=& \int \psi_{i}^{T*}(\mathbf{r})\,  \psi_{j}^T(\mathbf{r}) \, d^3r      \,.
\label{spinden}
\end{eqnarray}
These matrices
tell us about the inherent coherence of the spinor condensate and the thermal cloud. It
is convenient to consider the normalized spin density matrices 
$\tilde{\rho}_{ij} = \rho_{ij} / \sqrt{\rho_{ii} \rho_{jj}}$. The diagonal elements of
such matrices are equal to one and their off-diagonal elements determine the degree of 
coherence between various components of spinor condensate or the thermal cloud. 
When the off-diagonal element takes the value of the order of one (note that the maximal
value of $|\tilde{\rho}_{ij}|$ equals $1$) the coherence is strong whereas opposite 
means that the components are completely dephased.

\begin{figure}[thb]
\begin{center} \resizebox{3.in}{3.2in}
{\includegraphics{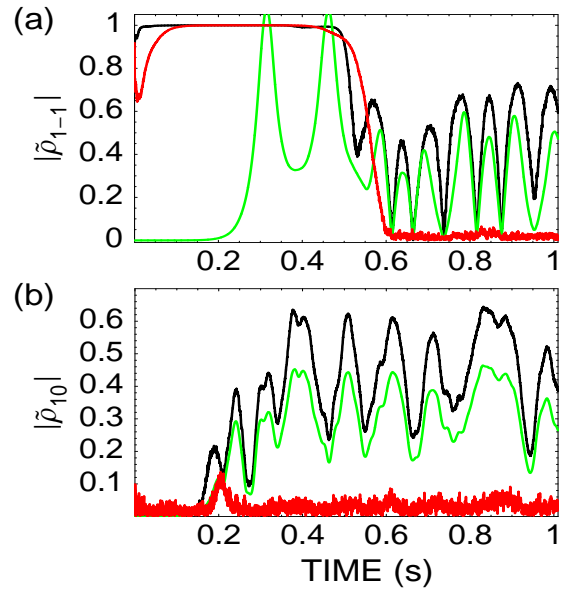}}
\caption{(color online). Spin density off-diagonal elements for 
(a) $(+1,-1)$ components ($c_2=-1.0\times10^{-3},\,c_0=0.22$, and $E_{kin}(0)=32\times10^3$) 
and (b) $(+1,0)$ components ($c_2=-2.29\times10^{-3},\,c_0=0.50$, and $E_{kin}(0)=
22\times10^3$) for the condensate (black line) and the thermal cloud (red line) as a 
function of time. The overlaps $|(\psi_{+1},\psi_{-1})|$ and $|(\psi_{+1},\psi_{0})|$ 
are plotted in green in (a) and (b), respectively.}
\label{sdoff}
\end{center}
\end{figure}

Coherence properties of $^{87}$Rb spinor condensate are displayed in Fig. \ref{sdoff},
where we show time dependence of spin density matrix elements corresponding to different
regimes of parameters according to the stability analysis. When the 
coupling constant $c_2$ is negative (ferromagnetic phase) and large enough to develop
the rotational instability (i.e., when modes $m=\pm1$ become unstable)
the strong coherence between the $+1$ and $-1$ spinor condensate
components builds up (upper frame). This coherence is maximal while the system is
approaching the thermal equilibrium and remains strong afterwards, however, in a way that it is
periodically lost and restored. Surprisingly, even thermal cloud in components $+1$ and $-1$
is fully coherent during the way of the system to its equilibrium but later on (after $0.6$\,s)
the thermal components get completely dephased. In this particular regime of parameters 
$(c_2,c_0)$ we observe only a development of coherence 
between $+1$ and $-1$ components. This can be understood by comparing the overlaps 
$|(\psi_{+1},\psi_{-1})|$ and $|(\psi_{+1},\psi_{0})|$ during the evolution -- the latter one 
turns out to be negligible. 

However, the situation changes when the strength of spin-changing collisions $c_2$ becomes
large enough (and still negative) to develop an instability leading to significant population
of $m_F=\pm1$ Zeeman states with nonrotating atoms (i.e., when the mode $m=0$ becomes 
unstable). In the case of Fig. \ref{sdoff}b the system reaches 
its equilibrium already after $0.2$\,s and then the overlap $|(\psi_{+1},\psi_{0})|$ gets even 
larger than the $|(\psi_{+1},\psi_{-1})|$ one and it results in a development of coherence 
between $+1$ and $0$ components. The same is true for the pair of $-1$ and $0$ states. 
Therefore, in this range of $(c_2,c_0)$ parameters all components of spinor condensates remain 
mutually coherent. Note that the thermal parts in $+1$ and $0$ states are always dephased.

Both frames show also the overlaps: $|(\psi_{+1},\psi_{-1})|$ in (a) and $|(\psi_{+1},\psi_{0})|$ 
in (b) case in green color. In the equilibrium the spin density off-diagonal elements follow 
the corresponding overlaps. It clearly indicates that the coherence is built 
as a result of many atomic collisions. They allow to establish the relative phase of the
components. In case (a) in the equilibrium the phase of the off-diagonal element 
$\tilde{\rho}_{1-1}$ becomes fixed within the intervals which correspond to the maximal
overlap but changes going from one maximum to the other. On the other hand, in (b) the
phase of $\tilde{\rho}_{10}$ gets constant just after the system reaches the thermal
equilibrium (i.e., after $0.2$\,s).

\begin{figure}[thb]
\begin{center} \resizebox{3.2in}{3.2in}
{\includegraphics{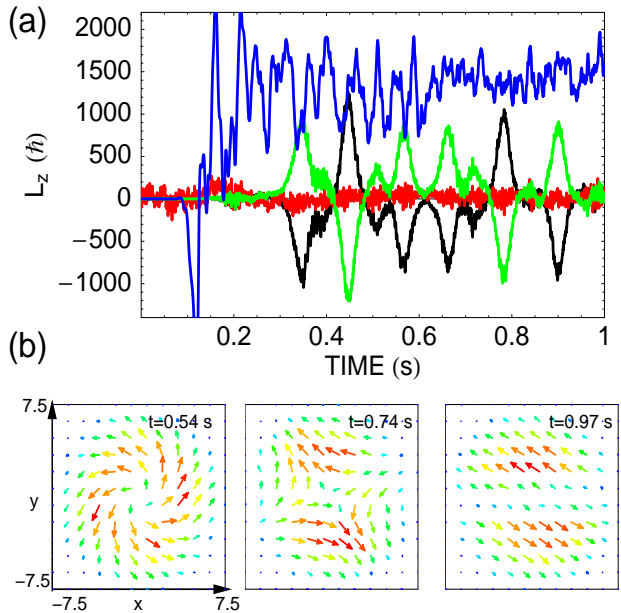}}
\caption{(color online). (a) Angular momentum of components $m_F=+1$ (black curve), 
$m_F=0$ (red curve), and $m_F=-1$ (green curve) as a function of time for parameters
$c_2=-1.0\times10^{-3}$, $c_0=0.22$, and $E_{kin}(0)=9\times10^3$. Blue curve shows the 
angular momentum of $m_F=+1$ state in the case when $c_0/c_2 \neq -216.4$ 
($c_2=-4.06\times10^{-3},\,c_0=0.25$).
(b) Spin textures at particular instants of time for parameters ($c_2,c_0$) as in the
first case of part (a). The length of the vector (as well as the color) 
represents the quantity $(<F_x>^2+<F_y>^2)^{1/2}$ and the size of each frame is 
$15 \times 15$ oscillatory units.}
\label{angmom}
\end{center}
\end{figure}

Another peculiar feature of spinor condensates revealed after the system reaches its 
equilibrium is breaking of the chiral symmetry (at zero temperature this phenomenon was 
already reported in \cite{chiral}). Since the modes $m=\pm1$ get unstable the atoms in 
components $+1$ and $-1$ start to rotate in opposite directions (the total orbital
angular momentum projection of colliding atoms is preserved by the contact interactions) 
on the onset of the equilibrium. No spontaneous rotation is observed when the system is 
approaching the equilibrium state. Fig. \ref{angmom}a (curves black, red, and green) shows 
the typical behavior of the angular momentum of all components when $c_0/c_2 = -216.4$
(this is the natural ratio for $F=1$ $^{87}$Rb atoms). In components $+1$ and $-1$ the 
atoms rotate alternatively clockwise and counterclockwise. This property changes when 
$c_0/c_2$ differs strongly from $-216.4$. In this case components rotate only in 
one direction permanently breaking the chiral symmetry (blue curve in Fig. \ref{angmom}a). 
Note, that unlike in \cite{chiral} this permanent rotation is obtained here in the 
conservative, hamiltonian dynamics.

It turns out that the periods the atoms in $m_F=+1$ or $m_F=-1$ states rotate in one 
direction (black and green lines in Fig. \ref{angmom}a) coincide with those when the 
off-diagonal spin density matrix element $|\tilde{\rho}_{1-1}|$ takes large values, i.e., 
when the mutual coherence between $m_F=+1$ and $m_F=-1$ components is built up. In these 
time intervals the system develops also highly regular spin textures. Examples are given 
in lower panel in Fig. \ref{angmom} where we plot the vector field composed of the average 
values of the total spin at particular instants of time. The first (from the left) frame 
shows the spiral galaxy-like pattern which persists within the interval $(0.52-0.62)$\,s. 
Later on a quadrupole-like texture is developed and lasts from $0.74$\,s to $0.81$\,s 
(the second frame). The third plot shows domain-like pattern. On the other hand, for the 
case when atoms rotate in one direction (blue curve) the system exhibits all the time the 
same structure which for the parameters considered is a spiral galaxy-like one.

In conclusion, we have shown that the initially randomly disturbed spinor condensate enters 
the equilibrium state in a step-like process. Counter intuitively, the dynamics
of the system becomes rich when it gets into the equilibrium. The atoms in components 
$m_F=+1$ and $m_F=-1$ spontaneously rotate in opposite directions and the magnetization 
shows the regular textures. Moreover, the system exhibits coherence properties dependent 
on the strength of the spin-changing collisions.
When such collisions are not strong enough but still allowing the transition to other
Zeeman states only spinor components $m_F=\pm 1$ develop the mutual coherence. We observe
that this coherence is periodically lost and restored. Simultaneously, the corresponding 
thermal parts get completely dephased. For larger values of coupling constants $c_2$ the
system enters a new regime where the coherence between components $m_F=0$ and $m_F=\pm 1$
is developed and becomes as strong as for $m_F=+1$ and $m_F=-1$ states.

\acknowledgments 
M.B., and M.G. acknowledge support by the Polish KBN Grant No. 1 P03B 051 30. 
K.G. thanks the Polish Ministry of Scientific Research Grant Quantum Information 
and Quantum Engineering No. PBZ-MIN-008/P03.
This paper (K.R.) was funded by Polish Government research funds for 2006-2009.

\end{document}